\documentclass[twocolumn,trackchanges]{aastex631}
\usepackage{amsmath,amssymb}
\usepackage{multirow}
\usepackage{float}
\usepackage[version=4]{mhchem}
\usepackage{tabularx}
\usepackage{threeparttable}
\usepackage{booktabs}
\usepackage{subfigure}
\usepackage{comment}
\usepackage{tablefootnote}
\begin{document}

\title{Rotational Spectra and Search for Aromatic Imines: 9-Iminofluorene and Benzophenone imine} 

\author[0009-0003-9416-5550]{Huanyu Ren}
\affiliation{Department of Chemistry, Massachusetts Institute of Technology, Cambridge, MA 02139, USA}

\author[0009-0005-1773-8460]{D. Archie Stewart}
\affiliation{Department of Chemistry, Massachusetts Institute of Technology, Cambridge, MA 02139, USA}

\author[0000-0002-0332-2641]{Gabi Wenzel}
\affiliation{Department of Chemistry, Massachusetts Institute of Technology, Cambridge, MA 02139, USA}
\affiliation{Center for Astrophysics \textbar{} Harvard \& Smithsonian, Cambridge, MA 02138, USA}

\author[0000-0001-8134-5681]{Thomas H. Speak}
\affiliation{Department of Chemistry, University of British Columbia, Vancouver, BC, Canada}

\author[0009-0008-1171-278X]{Martin S. Holdren}
\affiliation{Department of Chemistry, Massachusetts Institute of Technology, Cambridge, MA 02139, USA}

\author[0009-0002-6372-9926]{Reace H. J. Willis}
\affiliation{Department of Chemistry, University of British Columbia, Vancouver, BC, Canada}

\author[0000-0003-1254-4817]{Brett A. McGuire}
\affiliation{Department of Chemistry, Massachusetts Institute of Technology, Cambridge, MA 02139, USA}
\affiliation{National Radio Astronomy Observatory, Charlottesville, VA 22903, USA}

\correspondingauthor{Huanyu Ren, Brett A. McGuire}
\email{hyren@mit.edu, brettmc@mit.edu}

\begin{abstract}
Interstellar detections of several cyano derivatives of large polycyclic aromatic hydrocarbons (PAHs) have now been achieved, enabled by accurate laboratory measurements of their microwave rotational spectra. These results highlight the continued promise of other N-containing unsaturated PAHs, such as aromatic imines, as candidates for future laboratory studies and astronomical searches. In this work, we present broadband spectroscopic measurements of 9-iminofluorene and benzophenone imine in the 6-18\,GHz band. These measurements yield accurate rotational, centrifugal distortion, and $^{14}$N quadrupole coupling constants for both molecules. Using these experimentally-derived constants, we attempted a search for both molecules in the cold molecular cloud TMC-1 using observations from the Green Bank Telescope (GBT). Neither of these two ketimines was detected above the current noise level, establishing upper limits for their column densities of $5.1\times10^{12}\,\text{cm}^{-2}$ for 9-iminofluorene and $1.3\times10^{13}\,\text{cm}^{-2}$ for benzophenone imine. We also attempted a search for phenylmethanimine (both E/Z isomers) as the simplest aromatic aldimine, but neither was detected in TMC-1. To provide insight into these non-detections, we propose and evaluate different formation pathways using respective potential energy surfaces as determined by high-precision quantum chemical calculations. The result suggests the presence of an entrance barrier to forming the intermediate species, potentially explaining the low abundance. 
\end{abstract}

\keywords{astrochemistry -- interstellar medium -- molecular clouds -- microwave spectroscopy} 

\section{Introduction} \label{sec:intro}
Interstellar polycyclic aromatic hydrocarbons (PAHs) have recently been detected in surprisingly high abundance in the Taurus molecular cloud (TMC-1) and with a remarkable diversity in their size distribution. To date, the inventory contains 2-ring species, such as cyanonaphthalene \citep{cyanonaphthalene}, indene \citep{indeneB,indeneC}, and cyanoindene \citep{cyanoindene}; it also includes some larger PAHs, such as the 3-ring phenalene \citep{phenalene}, cyanoacenaphthylene \citep{cyanoacenaphthylene}, 4-ring cyanopyrene \citep{cyanopyrene,cyanopyrene2}, and the largest, 7-ring PAH cyanocoronene \citep{cyanocoronene}.

Despite the explosion in PAH detections in recent years, the puzzle of the PAH family in TMC-1 is still missing critical pieces. Fluorene is a special kind of 3-ring PAH with a five-membered ring between two six-membered rings, similar to detected acenaphthylene \citep{cyanoacenaphthylene}. Starting from indene, fluorene could be produced by the addition of a second six-membered ring onto its five-membered ring. As a result, it is reasonable to hypothesize that fluorene also exists in this environment in non-trivial abundance. However, accurate rotational spectroscopic parameters of fluorene have been measured by \citet{smallPAHs}, and no interstellar detection has been reported up to now, which might be due to its low permanent electric dipole moment (0.53 D). Since the intensity of the rotational lines is proportional to the square of the dipole moment, the signals from fluorene, even if present in similar abundances to other PAHs, may be challenging to detect.  The addition of a -CN group to molecules substantially increases their dipole moments, resulting in new, far stronger, and often more detectable signals. For example, the dipole moments of cyanofluorene isomers are \,$\sim$\,3--5 D. To help in the search for fluorene, \citet{cyanofluorene} measured rotational spectra and determined spectroscopic parameters for all five cyano derivatives of fluorene, but there have been no reports of any detection of them in space yet. This result was unexpected. It may indicate the presence of an unrecognized barrier to forming fluorene, or suggest that other, as-yet-undetected derivatives of fluorene that beyond the cyano series remain to be discovered.

Imines can be another potential class of derivatives of fluorene and other PAHs to explore in TMC-1. They are another interesting group of organic derivatives in space, considering that their \ce{C=N} bonds are very similar to \ce{C#N} bonds in nitriles. Several small, aliphatic imines have been detected in Sgr B2 and G+0.693-0.027: the simplest methanimine (\ce{CH2NH},  \citealp{methanimine1,methanimine2}); imines with a methyl group (\ce{CH3CHNH}, \citealt{ethanimine}), or a cyano group (E-NCCHNH, \citealp{Ecyanomethanimine}, and Z-NCCHNH, \citealp{Zcyanomethanimine}), or an acetylene group (\ce{HCCCHNH}, \citealp{HCCCHNH}). Some other relevant molecules detected include \ce{C=N} bonds, such as symmetrical carbodiimide (HNCNH, \citealp{HNCNH}), ketenimine (\ce{CH2CNH}, \citealp{ketenimine}), isocyanate (HNCO, \citealp{HNCO}), cyanomethanimine (\ce{H2CNCN}, \citealp{H2CNCN}), and even the imine radical methylene amidogen (\ce{H2CN}, \citealp{H2CN}) and its high-energy isomer aminocarbyne (\ce{H2NC}, \citealp{H2NC}). Although \ce{H2CN} might be the only one of these species detected in TMC-1 ($1.5 \times10^{11}\,\text{cm}^{-2}$ by \citealp{H2CN}, but this result has been questioned by \citealp{H2NC}), its presence nevertheless indicates the potential for imines in this astronomical source. Despite the significant reservoir of PAHs in TMC-1, there is a scarcity of studies exploring the potential detection of imines, notably aromatic imines, within this environment. This highlights a critical research gap and thus provides a strong motivation to investigate the presence of PAHs with imine functional groups, such as 9-iminofluorene and benzophenone imine, in TMC-1.

\begin{figure}[h!]

\begin{center}
\includegraphics[width=\columnwidth]{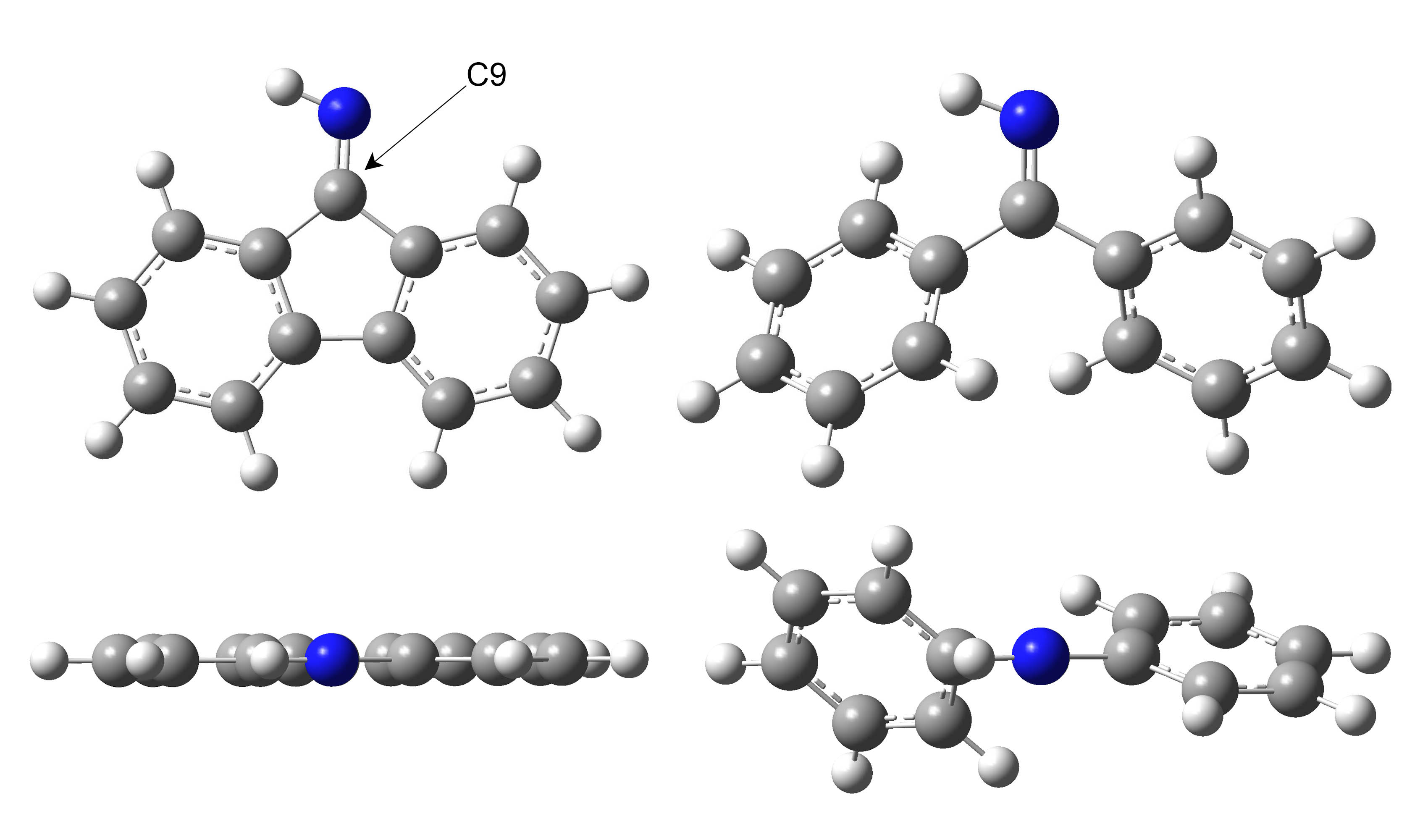}
\caption{
The different views of B3LYP-D3(BJ)/6-311++G(d,p) level optimized geometric structures for 9-iminofluorene (left) and benzophenone imine (right). The arrow shows C9 position of iminofluorene.}
\label{fig:ketones}
\end{center}
\end{figure}

There are several reasons to consider 9-iminofluorene as a potential interstellar imine-PAH. First, considering the carbon tetravalency principle, only the C9 position of fluorene (shown in Figure~\ref{fig:ketones}) on the five-membered ring can form a \ce{C=N} bond. Furthermore, this position can only achieve its highest degree of unsaturation as an imine, rather than a nitrile, due to the existing two C-C single bonds. In comparison, the C9 position is the hardest position to attach a cyano group for fluorene, because the 9-cyanofluorene has the highest energy among all cyano derivatives of fluorenes \citep{cyanofluorene}, which makes it thermodynamically much harder to form in cold astronomical environments. However, the formation of the \ce{C=N} bond, whose carbon atom is sp$^2$ hybridized, could conjugate two rings and lead to a lower energy. In comparison, 9-cyanofluorene has a sp$^3$ carbon that totally isolates two rings. Therefore, there are benefits for the formation of five-membered ring imines in space.

There are also practical reasons why these species are appealing targets in the laboratory.  Not all five-membered ring imine derivatives, such as iminocyclopentadiene and iminoindene, are stable enough for storage and measurement. 9-iminofluorene is characterized by two distinct and stable aromatic rings situated in the same plane, exhibiting weak conjugation. This structural arrangement is crucial as it prevents potential spontaneous polymerization driven by undesired Diels-Alder reactions \citep{DA}. Normally, for such a kind of \ce{$\alpha-\beta$} unsaturated imine, it is a reactive dienophile because of the electron-withdrawing group inside. However, the intrinsic stability derived from the conjugated aromatic framework makes 9-iminofluorene particularly robust, avoiding some side reactions like spontaneous polymerization. 

\begin{figure*}[t]
\begin{center}
\includegraphics[width=\columnwidth*2]{iminofluorene.png}
\caption{
The proposed two possible formation routes for 9-iminofluorene, while two important intermediates are benzophenone imine and 2-cyanobiphenyl.}
\label{fig:routes}
\end{center}
\end{figure*}

As well, the 9-iminofluorene is flanked by two symmetric groups, which effectively eliminates the possibility of cis-trans isomerism. This structural feature is crucial as it prevents the dilution of the already weak interstellar emission signal that would otherwise occur due to the presence of multiple isomers. Similarly, the search for 2-propanimine benefited from the absence of Z/E isomers \citep{2propanimine}. The challenges posed by isomeric mixtures are well-documented in astrochemical observations. For instance, the initial detection of cyanomethanimine involved only its E-isomer \citep{Ecyanomethanimine}, with the Z-isomer \citep{Zcyanomethanimine} being reported six years later. Asymmetric imines in space often exhibit complex isomer kinetics, typical of non-thermodynamic equilibrium conditions \citep{EZratio}.

Apart from iminofluorene, benzophenone imine is also a promising imine-PAH candidate for laboratory measurement and interstellar searches, not only due to its structural similarity to iminofluorene, but also because it may serve as a key intermediate in the formation of iminofluorene. As shown in Figure~\ref{fig:routes}, two potential gas-phase formation pathways for 9-iminofluorene are proposed, both initiated by the reaction between benzonitrile and phenyl radical. Depending on the site of radical attack, the reactions can yield two possible intermediates: benzophenone imine and 2-cyanobiphenyl. In order to do thorough interstellar searches of these species, accurate laboratory spectra and derived spectroscopic constants are required. Although laboratory rotational constants have been partly reported for 2-cyanobiphenyl \citep{cyanobiphenyl}, the centrifugal distortion constants and the frequencies of the measured spectral lines are not yet reported comprehensively. In contrast, benzophenone imine has never been studied by rotational spectroscopy. This lack of laboratory data motivates the present measurements of benzophenone imine, alongside 9-iminofluorene, to evaluate which, if any, formation pathway might be active under interstellar conditions.

\begin{figure*}
\begin{center}
\includegraphics[width=\columnwidth*2]{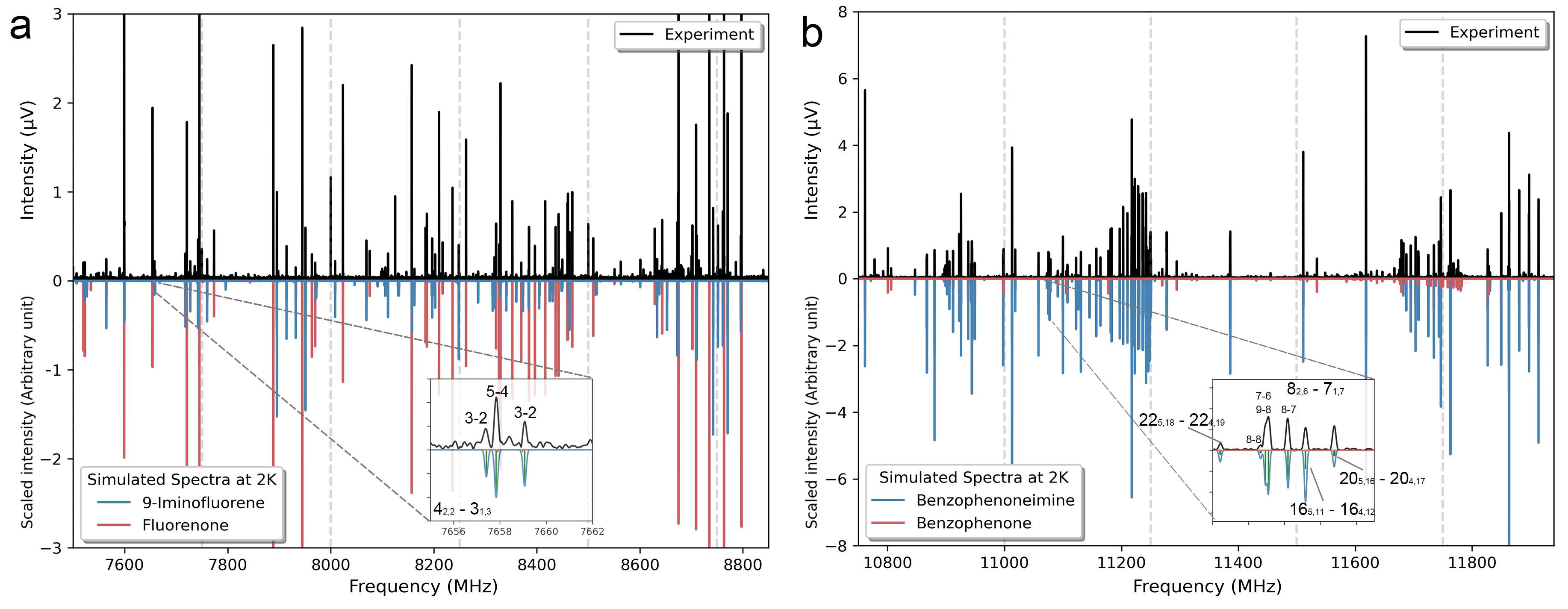}
\caption{
Parts of experimental spectra (black) overlaid with simulated fitting spectra (blue for imines, red for ketones) for (a) 9-Iminofluorene/Fluorenone and (b) Benzophenone imine/Benzophenone. The insets show zoomed-in views to highlight the resolved hyperfine splitting. For instance, in Figure a, the hyperfine components of $J_{K_a, K_c} = 4_{2,2} - 3_{1,3}$ transition from 9-iminofluorene is presented, with the changes in the fourth quantum number resulting from $^{14}$N quadrupole coupling clearly labeled. For the simulation lines, it has arbitrary units, and the intensity is scaled to simulate the experimental lines. The grey dotted line in the background marked the frequency as multiples of 250 MHz, and all strong lines at these frequencies could come from the clock as a reference in the spectrometer.}
\label{fig:spectra}
\end{center}
\end{figure*}
\section{Methods} \label{sec:methods}

\subsection{Microwave Spectroscopy}\label{subsec:spectroscopy}
A powder sample of 9-iminofluorene was purchased from Synthonix. The sample of benzophenone imine is liquid at room temperature and was purchased from Sigma-Aldrich. Both samples were used without further purification.  

Both samples' broadband rotational spectra (from 6 GHz to 18 GHz) were recorded using a compact chirped-pulse Fourier transform microwave (CP-FTMW) spectrometer \citep{Tweety}. To increase the vapor pressure of the solid sample 9-iminofluorene, the sample holder was heated to 110\,\textdegree C. The carrier gas (argon for 9-iminofluorene and neon for benzophenone imine)\footnote{Neon can provide improved signal quality but is prohibitively expensive compared to argon, so argon is used for most measurements.} was maintained at 1.28 kTorr and a mass-flow rate of 2.0 sccm to introduce the sample into a vacuum chamber held at the $10^{-5}$ Torr level. After entering the chamber, the sample vapor underwent supersonic expansion with the carrier gas, cooling the rotational temperature of the molecules to approximately 2 K. The nozzle generated the gas pulse at the rate of 3\,Hz with 600 $\mu$s width, and after 800 $\mu$s delay, each gas pulse was excited by 10 successive microwave chirp pulse cycles with 1 $\mu$s duration and 17 $\mu$s interval between excitation pulses. This 250 W high-power excitation chirp pulse, generated by an arbitrary waveform generator (AWG) and then amplified by a traveling wave tube (TWT) amplifier, was used to polarize the molecules. After each excitation chirp and the protecting delay, the free induction decay (FID) was collected for 10\,$\mu$s. Two million FIDs were averaged in the time domain. The resulting signal was processed with a Hann window function and then Fourier transformed into the frequency domain to yield the final spectrum for further analysis.

\subsection{Quantum Chemical Calculations}\label{subsec:computing}
To assign lines in the experimental spectra to the corresponding energy level transitions, quantum-chemical calculations were performed with the Gaussian 16 suite of programs (G16.C.02 version, \citealt{g16c02}) at the B3LYP-D3(BJ)/6-311++G(d,p) method and basis set \citep{B3LYP,6-311g,D3BJ} to optimize the geometry and determine rotational constants, which has been proven reliable to provide accurate geometric structure for similar size molecules\citep{Tweety,MRR}.  An anharmonic vibrational frequency calculation \citep{anharmonic, VPT2} was also applied to increase the accuracy of distortion constants. The computational results for rotational, centrifugal distortion, and hyperfine constants are reported in Tables~\ref{tab:constants}.

To evaluate the formation pathways of benzophenone imine and 9-iminofluorene via the reaction between benzonitrile and the phenyl radical, the potential energy surface (PES) was explored at the M06-2X-D3/def2TZVPP level \citep{cC3H2S}, and the energies were refined at the DLPNO-CCSD(T)/aug-cc-pVTZ level \citep{DLPNOCCSDT}. Specifically, the geometries of reactants, intermediates, transition states, and products were first optimized with the M06-2X functional \citep{M062X} with the D3 empirical dispersion correction \citep{D3} and the triple zeta def2-TZVPP basis set \citep{def2TZVPP1,def2TZVPP2}. Harmonic vibrational frequency calculations the stationary points and to obtain zero-point energy (ZPE) corrections using the Gaussian 16 program \citep{g16c02}. Subsequently, high-level single-point electronic energies (SPEs) were calculated with the Domain lone pair natural orbital approach to couple cluster with single double and perturbative triples (DLPNO-CCSD(T), \citealp{DLPNO}, tight PNO cut off) using restricted open shell Hartree-Fock reference energies (ROHF, \citealp{ROHF}), and applied aug-cc-pVTZ as main and auxiliary basis set \citep{augccpvtz,augccpvtz/c}, via the ORCA 6.0.1 program \citep{orca2020,orca6.0}. The reaction rate coefficients for the first steps proceeding via TS1 or TS2 were calculated using transition state theory (TST), with tunneling corrections through a one-dimensional Eckart barrier included \citep{tunneling}, as implemented in the KiSThelP \citep{KiSThelP} program.

\section{Results and Discussion} \label{sec:results}
\subsection{Laboratory Measurements}\label{subsec:measurements}
The least-squares fitting of the experimental microwave spectra was performed using the SPCAT/SPFIT suite \citep{spcat} to accurately determine the spectroscopic constants of 9-iminofluorene and benzophenone imine, including rotational constants, centrifugal distortion parameters, and $^{14}$N nuclear quadrupole coupling constants.

\begin{table*}
\centering
\footnotesize
\caption{Rotational and hyperfine constants of 9-iminofluorene and benzophenone imine in the A-reduced I$^r$ representation compared with the ones resulting from B3LYP-D3(BJ)/6-311++G(d,p) level anharmonic theoretical calculations.}
\begin{tabular}{l rr rr}
\toprule
\multirow{2}{*}{Parameter} & \multicolumn{2}{c}{9-Iminofluorene} & \multicolumn{2}{c}{Benzophenone imine} \\
\cmidrule(lr){2-3} \cmidrule(lr){4-5}
 & Computational & Experimental\textsuperscript{a} & Computational & Experimental\textsuperscript{a} \\
\midrule
$A$ (MHz)            & 1448.30  & 1445.22476(20)   & 1637.49  & 1639.18677(24) \\
$B$ (MHz)            & 585.76   & 585.653052(98)   & 417.45   & 415.845875(67) \\
$C$ (MHz)            & 417.07   & 416.898200(88)   & 362.83   & 361.651982(64)  \\
$\Delta_J$ (Hz)      & 3.73     & 4.16(19)         & 13.26    & 13.493(83) \\
$\Delta_{JK}$ (Hz)   & 4.57     & 3.83(12)         & -43.24   & -50.18(55) \\
$\Delta_K$ (Hz)      & 39.80    & 39.3(20)         & 462.93   & 457.2(44) \\
$\delta_J$ (Hz)      & 1.18     & 1.334(79)        & 1.86     & 1.890(29) \\
$\delta_K$ (Hz)      & 11.01    & 13.0(11)         & 58.85    & 57.6(23) \\
$\chi_{aa}$ (MHz)    & -2.32    & -2.179(10)       & -2.34    & -2.230(10) \\
$\chi_{bb}$ (MHz)    & -0.60    & -0.8827(39)      & -0.48    & -0.5722(81) \\

\\
$|\mu_a|$ (D)        & 1.3        & Observed          & 1.3        & Observed \\
$|\mu_b|$ (D)        & 2.0        & Observed          & 2.1        & Observed \\
$|\mu_c|$ (D)        & 0.0        & Not Observed      & 0.0      & Not Observed \\ \\
$N^\mathrm{fit}_\mathrm{lines}$ \textsuperscript{b}  & – & 357             & –          & 584 \\
$N^\mathrm{hyperfine}_\mathrm{lines}$ \textsuperscript{c} & – & 52         & –         & 133 \\
$\sigma_\mathrm{fit}$ (kHz) & – & 22.29              & –           & 15.61 \\
$(J, K_a)_\mathrm{max}$ & – & (21, 10)              & –            & (27, 11) \\
\bottomrule
\end{tabular}
\begin{tablenotes}
\item[a] a. Values in parentheses are 1$\sigma$ uncertainties in units of the last digit.
\item[b] b. The number of assigned experimental peaks. One peak can be assigned to several transitions due to limited resolution.
\item[c] c. The number of hyperfine resolved groups, in which transitions have the same $J,K_a,K_c$ but different experimental frequency.
\end{tablenotes}
\label{tab:constants}
\end{table*}

For 9-iminofluorene, 688 distinct rotational transitions were assigned to 350 experimental peaks, spanning quantum states up to $J_\mathrm{max} = 21$ and $K_a = 10$. Among these, 52 groups of transitions exhibited resolvable hyperfine structure, allowing for the precise determination of the $^{14}$N quadrupole coupling constants. The aromatic, coplanar polycyclic framework of 9-iminofluorene is highly rigid and stable, containing no freely rotatable covalent bonds or flexible segments. This rigidity results in smaller centrifugal distortion constants for iminofluorene. In contrast, due to the higher vapor pressure, the observed spectrum of benzophenone imine was more intense than that of 9-iminofluorene, yielding 455 assigned transitions, 100 groups of which showed resolved hyperfine structure. Figure~\ref{fig:spectra} displays representative parts of experimental spectra, including resolved hyperfine splitting and signals from impurity ketones. The increased number of transitions, especially those with high $J$ and $K$ values (up to $J_\mathrm{max} = 27$), facilitated more accurate determination of the centrifugal distortion constants $\delta_J$ and $\delta_K$. All resulting spectroscopic constants of these two molecules are summarized in Table~\ref{tab:constants}, based on a Watson $A$-reduction in the I$^r$ representation.

Because imines are highly susceptible to hydrolysis, during the spectroscopic measurement, 9-iminofluorene and benzophenone imine slowly reacted with the residual amount of water vapor in the carrier gas to produce 9-fluorenone and benzophenone, respectively. This side reaction is even more efficient under high-temperature conditions in the experiment. As a result, in the original data, lines from 9-fluorenone and benzophenone became stronger when the integration time increased. To identify the lines coming from these ketones, all ketone lines were also simulated from their published spectroscopic constants \citep{fluorenone}. 

\subsection{Astronomical Search toward TMC-1}\label{subsec:searching}

The interstellar search for 9-iminofluorene and benzophenone imine in TMC-1 was conducted based on the data from the 100\,m Robert C. Byrd Green Bank Telescope (GBT) as part of the GBT Observations of TMC-1: Hunting Aromatic Molecules (GOTHAM) large program \citep{GOTHAM}. GOTHAM underwent several data reductions (DR2, \citealt{cyanonaphthalene}; DR3, \citealt{Heterocycles}; DR4, \citealt{cyanoindene, Ecyanobutadiene}), and the latest fifth data reduction (DR5.3, \citealt{DR5.3}) was applied in this study. It not only includes some updates on radio-frequency interference removal, atmospheric opacity corrections, and baseline removal, but also considers the frequency-dependent beam efficiency directly. The pointing position was toward the ``cyanopolyyne peak'' in TMC-1 at $\alpha_\mathrm{J2000}$ = 04\fh41\fm42.5\fs, $\delta_\mathrm{J2000}$ = +25\arcdeg41\arcmin26.8\arcsec. The survey covers 29 GHz in bandwidth (near frequency continuous between 3.9 GHz and 36.4 GHz), with a spectral resolution of 1.4 kHz. This is an ideal range for the search for molecules with more than 10 heavy atoms, because heavier molecules have larger moments of inertia, the peak of the Boltzmann distribution of occupied transitions for TMC-1 conditions temperature shifts to lower frequencies, often within this bandwidth. Because it was the combination of different observations with several receivers and varied integration time, the RMS noise level fluctuates between 2--20 mK in different spectrum segments. 

The rotational spectra of 9-iminofluorene and benzophenone imine were simulated and searched at typical conditions for TMC-1 \((T = 7\,\mathrm{K},\ v = 5.8\,\mathrm{km\,s^{-1}})\). After comparing with the GOTHAM DR5.3 data, no individual transitions were identified for either of the two species at the current level of sensitivity. Using the transition that would provide the strongest constraint on column density (i.e. the line that would be the highest signal-to-noise-ratio if present), $3\sigma$ upper limits to the column density of each molecule were derived to be $5.1\times10^{12}\,\text{cm}^{-2}$ for 9-iminofluorene (at 18759.926 MHz) and $1.3\times10^{13}\,\text{cm}^{-2}$ for benzophenone imine (at 19197.428 MHz). To further explore their existence due to their complex spectra, which will dilute the intensity of the single line, we performed a velocity stack and matched filtering analysis on both species for the 150 strongest lines \citep{stacking,cyanonaphthalene}. No signal was detected in the stacked analysis either, which suggests the true upper limit may be even 1-2 orders of magnitude lower \citep{Heterocycles}.

\begin{figure}[h!]
\begin{center}
\includegraphics[width=\columnwidth]{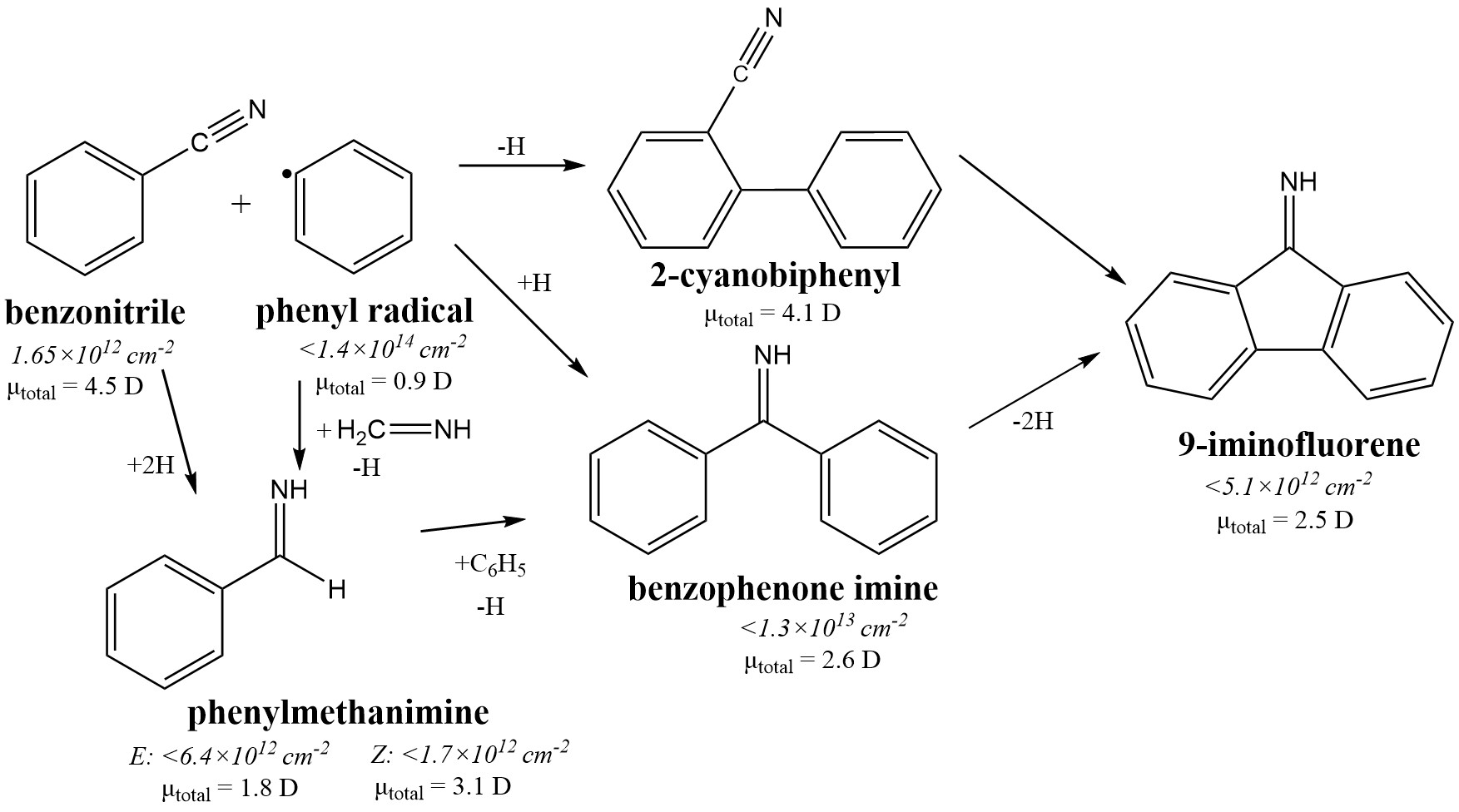}
\caption{Possible pathways to form 9-iminofluorene from benzonitrile that are discussed in this work. Apart from the column density of benzonitrile reported by \citealp{DR5.3}, all other upper limits are reported in this work using the same derivation method.}
\label{fig:pathways}
\end{center}
\end{figure}

\subsection{Insights to Formation Mechanism}\label{subsec:discussion}
The search for both benzophenone imine and 9-iminofluorene in TMC-1 could provide new insights into the proposed mechanism illustrated in Figure~\ref{fig:routes}, as well as other related radical–neutral reaction pathways. 
Since different sequences of bond formation can converge on the same final product, Figure~\ref{fig:pathways} summarizes several plausible routes by which 9-iminofluorene could originate from benzonitrile—the closest detected precursor to both imine species. While these routes do not exhaust all potential formation mechanisms, they were selected for their relative simplicity and their reliance on intermediates either already observed or considered plausible in TMC-1. For example, pathways involving species such as fluorene or biphenyl are not discussed here, as these have not yet been detected in this source.

The lack of sufficient column density for two intermediates -- interstellar benzophenone imine, studied here, plus the possible absence of 2-cyanobiphenyl \citep{cyanobiphenyl} -- suggests an explanation for why 9-iminofluorene cannot be detected at the current noise level. As for the concentration of two reactants at the beginning, benzonitrile and phenyl radical, even though benzonitrile has been unambiguously detected in TMC-1, its column density ($1.65 \times10^{12}\,\text{cm}^{-2}$, \citealp{DR5.3}) is still relatively low, lower than the upper limits of polycyclic products measured in this work. Although the phenyl radical is widely considered as a potential building block for larger PAHs \citep{phenylradical}, it has not yet been identified in space, which could be an indication of low abundance or it could be due to its low dipole moment (0.87\,D, \citealt{phenyl}) resulting in very weak potential signals. From the experimentally measured spectrum of phenyl radical \citep{phenyl}, we can derive the upper limit of it towards TMC-1 is $1.4\times10^{14}\,\text{cm}^{-2}$.

The two PAH imines studied here have very different structures compared to previously detected imines:  methanimine (\ce{CH2NH}), ethanimine (\ce{CH3CHNH}), cyanomethanimine (NCCHNH), and propargylimine (\ce{HCCCHNH}) all have hydrogen on imine carbons and their corresponding nitrile species. The hydrogenation reactions of nitriles have been proposed as possible formation routes for these detected imines \citep{HCCCHNH}. Both HCN and \ce{CH2NH} ices have been shown to yield \ce{CH3NH2} upon exposure to atomic hydrogen \citep{hydrogenation}, providing support for the hypothesis of ice-phase nitrile hydrogenation. However, if the hydrogenation of nitriles is indeed the dominant formation pathway for interstellar imines, then ketimines—such as benzophenone imine and iminofluorene—would be expected to have relatively low column densities, as they cannot form through this route. The absence of such a hydrogenation-based pathway may also explain the non-detection of 2-propanimine—the simplest ketimine—in sources such as Sgr B2(N1) and IRAS 16293–2422, as reported in previous studies \citep{2propanimine}.

To further examine the role of nitrile hydrogenation reactions for imine-PAHs, we also searched for both the Z/E isomers of phenylmethanimine in TMC-1, because it is the simplest aromatic aldimine, and has already been spectroscopically characterized in the laboratory \citep{phenylmethanimine}. However, neither isomer of phenylmethanimine was detected, with derived upper limits of $8.18 \times10^{12}\,\text{cm}^{-2}$ for E isomer and $9.52 \times10^{11}\,\text{cm}^{-2}$ for Z isomer. This result could potentially challenge the assumption of nitriles' hydrogenation as the dominant pathway to form imines, but the upper limit cannot be strong evidence. Considering interstellar imines are usually not more abundant than their corresponding nitriles (for example, $\ce{HC3HNH}/\ce{HC3N}\approx0.03$ towards G+0.6930.027, \citealp{HCCCHNH}), this may partly explain the failure of this detection, because the column density of benzonitrile is even lower than the upper limit of phenylmethanimine. While methanimine could also be an important intermediate for other complex imines \citep{Methaniminepath}, the missing phenylmethanimine suggests the formation route from methanimine to phenylmethanimine, and then benzophenone imine, may be inefficient.

\begin{figure*}[t!]
\begin{center}
\includegraphics[width=\textwidth]{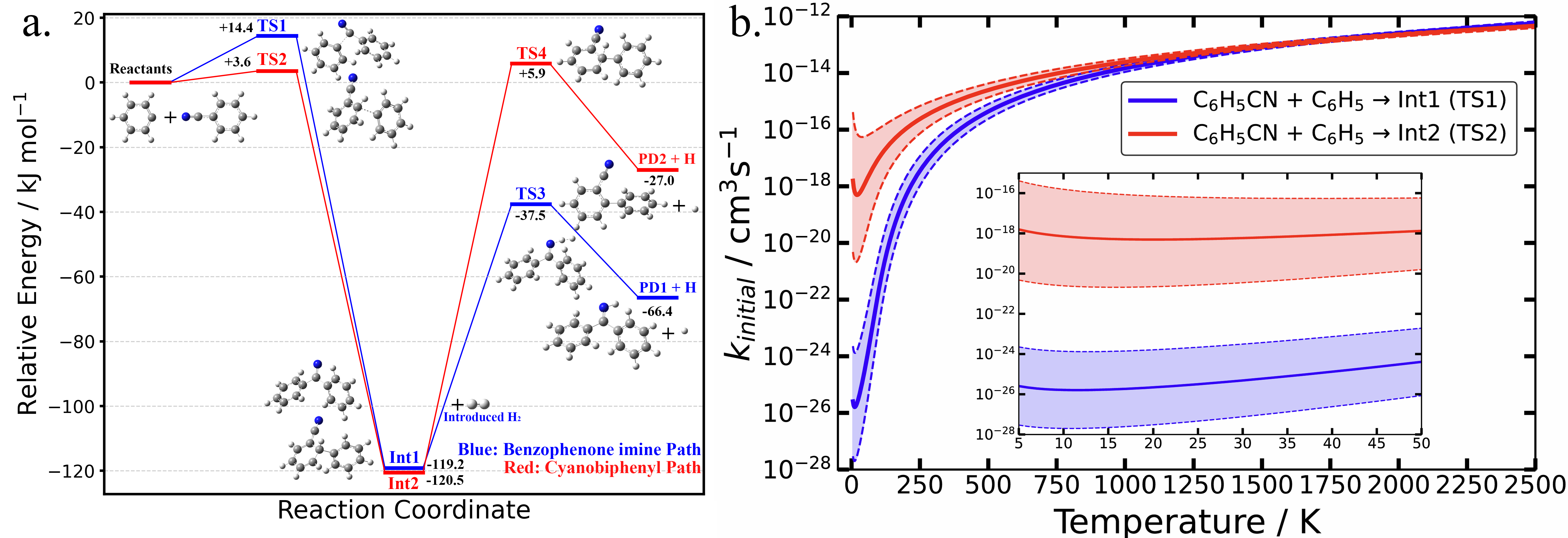}
\caption{(a) The calculated energy (relative to the separated reactants, including  \ce{H2}) for transition states and intermediates involved in the formation routes for benzophenone imine (blue) and cyanobiphenyl (red) under DLPNO-CCSD(T)/aug-cc-pVTZ//M06-2X-D3/def2-TZVPP level. (b) Calculated reaction rate coefficients obtained from transition state theory with tunneling corrections included using the Kinetic and Statistical Thermodynamical Package (KiSThelP,  \citealp{KiSThelP}). The shaded regions represent the propagated uncertainty in the barrier heights of TS1 and TS2 (±\(2.5\,\mathrm{kJ\,mol^{-1}}\)), with \(+2.5\,\mathrm{kJ\,mol^{-1}}\) corresponding to the lower red and blue dashed lines, and \(-2.5\,\mathrm{kJ\,mol^{-1}}\) to the upper dashed lines.}
\label{fig:pes}
\end{center}
\end{figure*}

No small imines have been detected in TMC‑1 apart from tentatively detected radical \ce{H2CN} \citep{H2CN}. If \ce{H2CN} does exist, it could potentially react with benzene or phenyl radical to form phenylmethanimine, which might then proceed to benzophenone imine through a second collision. However, since phenylmethanimine was not detected in our observations, a direct single-step reaction between benzonitrile and phenyl radical appears to be a more feasible route. This reaction could yield benzophenone imine or cyanobiphenyl, followed by intramolecular cyclization to form iminofluorene (Figure~\ref{fig:routes}). To evaluate the feasibility of these pathways, we computed the potential energy surfaces (PES) for both the formation of benzophenone imine and cyanobiphenyl as two precursors for iminofluorene. The relative energies of transition states and intermediates are shown in Figure~\ref{fig:pes}.

As concluded by \citet{gasphasekinetics}, the extremely low temperature and density of TMC-1 require binary gas-phase neutral–neutral reactions to be both exothermic and to involve at least one highly reactive species, such as the phenyl radical discussed here. Quantum chemical calculations show that both reaction pathways considered here are thermodynamically favorable,  with the formation of 2-cyanobiphenyl and benzophenone imine being exothermic by \(-27.0\,\mathrm{kJ\,mol^{-1}}\) and \(-66.4\,\mathrm{kJ\,mol^{-1}}\), respectively. This suggests that the benzophenone imine pathway is more favorable in terms of reaction enthalpy.

However, kinetic considerations are equally, if not more important: the activation energy associated with transition states and intermediates plays a critical role in determining whether a reaction can proceed in cold molecular clouds, regardless of its thermodynamic favorability. The lack of any additional thermal energy only allows for barrierless entrance and exit channels, and for submerged barriers within the reaction path \citep{gasphasekinetics, addCN}.  The calculations here reveal that the first step of both proposed reaction pathways have significant energy barriers. Specifically, for the benzophenone imine path, even though the second step a submerged barrier (\(-37.5\,\mathrm{kJ\,mol^{-1}}\)), its first step requires 
\(14.4\,\mathrm{kJ\,mol^{-1}}\) positive activation energy (\(\mathrm{\approx 1730\,\mathrm{K}}\) in temperature units), which is nearly impossible to overcome in the \(7\,\mathrm{K}\) TMC-1 environment.  If there is no other viable pathway to forming benzophenone, then this potentially explains the non-detection of benzophenone imine in observations. Similarly, the cyanobiphenyl pathway also involves transition states with positive relative energies, suggesting that it is also unlikely to occur efficiently under cold molecular cloud conditions.

While considering tunneling corrections, the highest TST reaction rate coefficients in Figure~\ref{fig:pes} under TMC-1 conditions $(\approx 7\,K)$ are around \(10^{-16}\,\mathrm{cm^{3}\,s^{-1}}\) for the cyanobiphenyl pathway and \(10^{-24}\,\mathrm{cm^{3}\,s^{-1}}\) for the benzophenone imine pathway. To estimate the time required for product densities to reach their observed upper limits, we evaluate the most optimistic scenario—assuming only the rate-limiting first step occurs. For the benzophenone imine pathway, this assumption is reasonable since the second step has a submerged barrier. Although the cyanobiphenyl pathway exhibits a much higher second-step barrier, if the first step itself is inaccessible, the entire pathway can be ruled out. Consequently, our timescale estimates here consider only the initial steps of both reactions.

We also assume that the column density of the phenyl radical reaches an upper limit of \(N_{max}(\ce{C6H5})\approx\,10^{14}\,\mathrm{cm^{-2}}\) to make the reaction faster, which is two orders of magnitude higher than that of benzonitrile \(N(\ce{C6H5CN})\approx\,10^{12}\,\mathrm{cm^{-2}}\) \citep{DR5.3}. Under this assumption, the bimolecular reaction can be treated as pseudo–first-order, and the reaction time required to reach a given product density is calculated using the following formula:
\begin{equation}
t \;=\; \frac{1}{k \, A_0} \, \ln\!\left(\frac{[B] + [AB]}{[B]}\right),
\label{eq:pseudo-first-order}
\end{equation}
where A, B, and AB denote the phenyl radical, the benzonitrile, the bimolecular reaction product.

To convert the column density (N) to the volume density (n), we adopt \ce{H2} in TMC-1 as reference, with \(n(\ce{H2})\approx\,10^{4}\,\mathrm{cm^{-3}}\), and \(N(\ce{H2})\approx\,10^{22}\,\mathrm{cm^{-2}}\) \citep{n(H2),n(H2)2}. This yields an upper limit for the phenyl radical volume density of \(A_0 = n_{max}(\ce{C6H5})\approx\,10^{-4}\,\mathrm{cm^{-3}}\), corresponding to \(N_{max}(\ce{C6H5})\approx\,10^{14}\,\mathrm{cm^{-2}}\). To estimate the benzophenone imine path, substituting [AB]\,=\(N_{max}(\ce{(C6H5)2C=NH})\approx\,10^{13}\,\mathrm{cm^{-2}}\), [B]\,=\(N(\ce{C6H5CN})\approx\,10^{12}\,\mathrm{cm^{-2}}\), and k\,=\(10^{-24}\,\mathrm{cm^{3}\,s^{-1}}\) into Eq.~\ref{eq:pseudo-first-order} gives $t=4.5 \times10^{22}\,\text{yr}$. Even for the faster cyanobiphenyl path with k\,=\(10^{-16}\,\mathrm{cm^{3}\,s^{-1}}\), the resulting timescale is still $t=4.5 \times10^{14}\,\text{yr}$. Both values far exceed the cosmic age ($\approx1.38 \times10^{10}\,\text{yr}$).
 
These extremely long reaction timescales suggest that the substantial energy barriers associated with both the benzophenone imine and cyanobiphenyl formation routes likely inhibit the efficient formation of 9-iminofluorene under TMC-1 condition, which might otherwise be produced via subsequent intramolecular cyclization.

\section{Conclusion} \label{sec:conclusion}
In this work, we measured the rotational spectra of two imine-PAHs -- benzophenone imine and 9-iminofluorene -- for the first time in the 6–18\,GHz region and derived their spectroscopic constants. These constants enabled a targeted search in TMC-1 using GOTHAM data. No signals were found in either single-line or stacking analyses, but the experimental constants can still guide the further search for these PAH imines in other astronomical sources. The hydrogenation pathway to produce aromatic imines was evaluated by searching for both isomers of phenylmethanimine without any signals. We calculated part of the PES for two proposed pathways to 9-iminofluorene, and both show positive barriers. The rate coefficients derived from these barriers point to strong kinetic inhibition, which may help explain why neither benzophenone imine nor cyanobiphenyl, nor consequently 9-iminofluorene itself, is observed.
\section{Data Availability} \label{sec:data}
The original spectroscopic files of two molecules, including lin files from experimental measurement, var files for fitted spectroscopic parameters, int files for simulation parameters, and cat files for observational searching, are all available on Zenodo under an open-source Creative Commons Attribution license: 
\dataset[doi:10.5281/zenodo.17611480]{https://doi.org/10.5281/zenodo.17611480}.

\begin{acknowledgements}

We gratefully acknowledge support from NSF grants AST-1908576, AST-2205126, and AST-2307137 and Schmidt Family Futures. G.W. and B.A.M. acknowledge the support of the Arnold and Mabel Beckman Foundation Beckman Young Investigator Award. The National Radio Astronomy Observatory is a facility of the National Science Foundation operated under cooperative agreement by Associated Universities, Inc. The Green Bank Observatory is a facility of the National Science Foundation operated under cooperative agreement by Associated Universities, Inc.  T.H.S. acknowledges the support of the Canadian Space Agency (CSA) through grant 24AO3UBC14. 

\textbf{Statement of Efforts.}  All authors contributed to the writing and editing of the manuscript.
\end{acknowledgements}

\bibliography{references}

@article{spcat,
  title = {The Fitting and Prediction of Vibration-Rotation Spectra with Spin Interactions},
  author = {Pickett, Herbert M.},
  year = {1991},
  journal = {Journal of Molecular Spectroscopy},
  volume = {148},
  number = {2},
  pages = {371--377},
  issn = {0022-2852},
  doi = {10.1016/0022-2852(91)90393-O},
}

@article{fluorenone,
  title = {Newly Assigned Microwave Transitions and a Global Analysis of the Combined Microwave/Millimeter Wave Rotational Spectra of 9-Fluorenone and Benzophenone},
  author = {West, Channing and Sedo, Galen and Van Wijngaarden, Jennifer},
  year = {2017},
  month = may,
  journal = {Journal of Molecular Spectroscopy},
  volume = {335},
  pages = {43--48},
  issn = {00222852},
  doi = {10.1016/j.jms.2016.12.002},
  urldate = {2024-12-24},
}

@article{Tweety,
  title = {Laboratory {{Rotational Spectra}} of {{Cyanocyclohexane}} and {{Its Siblings}} (1- and 4-{{Cyanocyclohexene}}) {{Using}} a {{Compact CP-FTMW Spectrometer}} for {{Interstellar Detection}}},
  author = {Wenzel, Gabi and Holdren, Martin S. and Stewart, D. Archie and Toru Shay, Hannah and Byrne, Alex N. and Xue, Ci and McGuire, Brett A.},
  year = {2025},
  month = may,
  journal = {The Journal of Physical Chemistry A},
  volume = {129},
  number = {18},
  pages = {3986--4001},
  issn = {1089-5639, 1520-5215},
  doi = {10.1021/acs.jpca.5c00980},
  urldate = {2025-05-14},
}

@article{cyanonaphthalene,
  title = {Detection of Two Interstellar Polycyclic Aromatic Hydrocarbons via Spectral Matched Filtering},
  author = {McGuire, Brett A. and Loomis, Ryan A. and Burkhardt, Andrew M. and Lee, Kin Long Kelvin and Shingledecker, Christopher N. and Charnley, Steven B. and Cooke, Ilsa R. and Cordiner, Martin A. and Herbst, Eric and Kalenskii, Sergei and Siebert, Mark A. and Willis, Eric R. and Xue, Ci and Remijan, Anthony J. and McCarthy, Michael C.},
  year = {2021},
  month = mar,
  journal = {Science},
  volume = {371},
  number = {6535},
  pages = {1265--1269},
  issn = {0036-8075, 1095-9203},
  doi = {10.1126/science.abb7535},
  urldate = {2025-04-19},
}

@article{indeneB,
  title = {Discovery of the {{Pure Polycyclic Aromatic Hydrocarbon Indene}} (c-{{C9H8}}) with {{GOTHAM Observations}} of {{TMC-1}}},
  author = {Burkhardt, Andrew M. and Long Kelvin Lee, Kin and Bryan Changala, P. and Shingledecker, Christopher N. and Cooke, Ilsa R. and Loomis, Ryan A. and Wei, Hongji and Charnley, Steven B. and Herbst, Eric and McCarthy, Michael C. and McGuire, Brett A.},
  year = {2021},
  month = jun,
  journal = {The Astrophysical Journal Letters},
  volume = {913},
  number = {2},
  pages = {L18},
  issn = {2041-8205, 2041-8213},
  doi = {10.3847/2041-8213/abfd3a},
  urldate = {2025-04-17},
}

@article{indeneC,
  title = {Pure Hydrocarbon Cycles in {{TMC-1}}: {{Discovery}} of Ethynyl Cyclopropenylidene, Cyclopentadiene, and Indene},
  shorttitle = {Pure Hydrocarbon Cycles in {{TMC-1}}},
  author = {Cernicharo, J. and Ag{\'u}ndez, M. and Cabezas, C. and Tercero, B. and Marcelino, N. and Pardo, J. R. and De Vicente, P.},
  year = {2021},
  month = may,
  journal = {Astronomy \& Astrophysics},
  volume = {649},
  pages = {L15},
  issn = {0004-6361, 1432-0746},
  doi = {10.1051/0004-6361/202141156},
  urldate = {2025-04-17},
}

@article{cyanoindene,
  title = {Discovery of {{Interstellar}} 2-{{Cyanoindene}} (2-{{C}}{\textsubscript{9}} {{H}}{\textsubscript{7}} {{CN}}) in {{GOTHAM Observations}} of {{TMC-1}}},
  author = {Sita, Madelyn L. and Changala, P. Bryan and Xue, Ci and Burkhardt, Andrew M. and Shingledecker, Christopher N. and Kelvin Lee, Kin Long and Loomis, Ryan A. and Momjian, Emmanuel and Siebert, Mark A. and Gupta, Divita and Herbst, Eric and Remijan, Anthony J. and McCarthy, Michael C. and Cooke, Ilsa R. and McGuire, Brett A.},
  year = {2022},
  month = oct,
  journal = {The Astrophysical Journal Letters},
  volume = {938},
  number = {2},
  pages = {L12},
  issn = {2041-8205, 2041-8213},
  doi = {10.3847/2041-8213/ac92f4},
  urldate = {2025-01-13},
}

@article{cyanoacenaphthylene,
  title = {Discovery of Two Cyano Derivatives of Acenaphthylene ({{C}}{\textsubscript{12}} {{H}}{\textsubscript{8}} ) in {{TMC-1}} with the {{QUIJOTE}} Line Survey},
  author = {Cernicharo, J. and Cabezas, C. and Fuentetaja, R. and Ag{\'u}ndez, M. and Tercero, B. and Janeiro, J. and Juanes, M. and Kaiser, R. I. and Endo, Y. and Steber, A. L. and P{\'e}rez, D. and P{\'e}rez, C. and Lesarri, A. and Marcelino, N. and De Vicente, P.},
  year = {2024},
  month = oct,
  journal = {Astronomy \& Astrophysics},
  volume = {690},
  pages = {L13},
  issn = {0004-6361, 1432-0746},
  doi = {10.1051/0004-6361/202452196},
  urldate = {2025-04-17},
}

@article{cyanopyrene2,
  title = {Detections of Interstellar Aromatic Nitriles 2-Cyanopyrene and 4-Cyanopyrene in {{TMC-1}}},
  author = {Wenzel, Gabi and Speak, Thomas H. and Changala, P. Bryan and Willis, Reace H. J. and Burkhardt, Andrew M. and Zhang, Shuo and Bergin, Edwin A. and Byrne, Alex N. and Charnley, Steven B. and Fried, Zachary T. P. and Gupta, Harshal and Herbst, Eric and Holdren, Martin S. and Lipnicky, Andrew and Loomis, Ryan A. and Shingledecker, Christopher N. and Xue, Ci and Remijan, Anthony J. and Wendlandt, Alison E. and McCarthy, Michael C. and Cooke, Ilsa R. and McGuire, Brett A.},
  year = 2025,
  month = feb,
  journal = {Nature Astronomy},
  volume = {9},
  number = {2},
  pages = {262--270},
  publisher = {Nature Publishing Group},
  issn = {2397-3366},
  doi = {10.1038/s41550-024-02410-9},
  urldate = {2025-11-20},
  copyright = {2024 The Author(s)},
  langid = {english}
}

@article{cyanocoronene,
  title = {Discovery of the {{Seven-ring Polycyclic Aromatic Hydrocarbon Cyanocoronene}} ({{C24H11CN}}) in {{GOTHAM Observations}} of {{TMC-1}}},
  author = {Wenzel, Gabi and Gong, Siyuan and Xue, Ci and Changala, P. Bryan and Holdren, Martin S. and Speak, Thomas H. and Stewart, D. Archie and Fried, Zachary T. P. and Willis, Reace H. J. and Bergin, Edwin A. and Burkhardt, Andrew M. and Byrne, Alex N. and Charnley, Steven B. and Lipnicky, Andrew and Loomis, Ryan A. and Shingledecker, Christopher N. and Cooke, Ilsa R. and McCarthy, Michael C. and Remijan, Anthony J. and Wendlandt, Alison E. and McGuire, Brett A.},
  year = {2025},
  month = apr,
  journal = {The Astrophysical Journal Letters},
  shortjournal = {ApJL},
  volume = {984},
  number = {1},
  pages = {L36},
  publisher = {The American Astronomical Society},
  issn = {2041-8205},
  doi = {10.3847/2041-8213/adc911},
  url = {https://dx.doi.org/10.3847/2041-8213/adc911},
  urldate = {2025-05-27},
}

@article{cyanopyrene,
  title = {Detection of Interstellar 1-Cyanopyrene: {{A}} Four-Ring Polycyclic Aromatic Hydrocarbon},
  shorttitle = {Detection of Interstellar 1-Cyanopyrene},
  author = {Wenzel, Gabi and Cooke, Ilsa R. and Changala, P. Bryan and Bergin, Edwin A. and Zhang, Shuo and Burkhardt, Andrew M. and Byrne, Alex N. and Charnley, Steven B. and Cordiner, Martin A. and Duffy, Miya and Fried, Zachary T. P. and Gupta, Harshal and Holdren, Martin S. and Lipnicky, Andrew and Loomis, Ryan A. and Shay, Hannah Toru and Shingledecker, Christopher N. and Siebert, Mark A. and Stewart, D. Archie and Willis, Reace H. J. and Xue, Ci and Remijan, Anthony J. and Wendlandt, Alison E. and McCarthy, Michael C. and McGuire, Brett A.},
  year = {2024},
  month = nov,
  journal = {Science},
  volume = {386},
  number = {6723},
  pages = {810--813},
  publisher = {American Association for the Advancement of Science},
  doi = {10.1126/science.adq6391},
  urldate = {2024-12-24},
}

@article{cyanofluorene,
  title = {Rotational Spectra of Five Cyano Derivatives of Fluorene},
  author = {Cabezas, Carlos and Janeiro, Jesús and Steber, Amanda L. and Pérez, Dolores and Bermúdez, Celina and Guitián, Enrique and Lesarri, Alberto and Cernicharo, José},
  year = {2024},
  journal = {Physical Chemistry Chemical Physics},
  shortjournal = {Phys. Chem. Chem. Phys.},
  volume = {26},
  number = {36},
  pages = {23703--23709},
  issn = {1463-9076, 1463-9084},
  doi = {10.1039/D4CP01924E},
  url = {https://xlink.rsc.org/?DOI=D4CP01924E},
  urldate = {2025-01-06},
}

@article{methanimine1,
  title = {Discovery of Interstellar Methanimine (Formaldimine)},
  author = {Godfrey, P. D. and Brown, R. D. and Robinson, B. J. and Sinclair, M. W.},
  year = {1973},
  month = feb,
  journal = {Astrophysical Letters},
  volume = {13},
  pages = {119}
}

@article{methanimine2,
  title = {Discovery of Methanimine ({{CH}}{\textsubscript{2}} {{NH}}) Megamasers toward Compact Obscured Galaxy Nuclei},
  author = {Gorski, M. D. and Aalto, S. and Mangum, J. and Momjian, E. and Black, J. H. and Falstad, N. and Gullberg, B. and K{\"o}nig, S. and Onishi, K. and Sato, M. and Stanley, F.},
  year = {2021},
  month = oct,
  journal = {Astronomy \& Astrophysics},
  volume = {654},
  pages = {A110},
  issn = {0004-6361, 1432-0746},
  doi = {10.1051/0004-6361/202141633},
  urldate = {2025-05-21},
}

@article{ethanimine,
  title = {{{THE DETECTION OF INTERSTELLAR ETHANIMINE}} ({{CH}}{\textsubscript{3}} {{CHNH}}) {{FROM OBSERVATIONS TAKEN DURING THE GBT PRIMOS SURVEY}}},
  author = {Loomis, Ryan A. and Zaleski, Daniel P. and Steber, Amanda L. and Neill, Justin L. and Muckle, Matthew T. and Harris, Brent J. and Hollis, Jan M. and Jewell, Philip R. and Lattanzi, Valerio and Lovas, Frank J. and Martinez, Oscar and McCarthy, Michael C. and Remijan, Anthony J. and Pate, Brooks H. and Corby, Joanna F.},
  year = {2013},
  month = feb,
  journal = {The Astrophysical Journal},
  volume = {765},
  number = {1},
  pages = {L9},
  issn = {2041-8205, 2041-8213},
  doi = {10.1088/2041-8205/765/1/L9},
  urldate = {2025-05-21},
}

@article{Ecyanomethanimine,
  title = {{{DETECTION OF E-CYANOMETHANIMINE TOWARD SAGITTARIUS B2}}({{N}}) {{IN THE GREEN BANK TELESCOPE PRIMOS SURVEY}}},
  author = {Zaleski, Daniel P. and Seifert, Nathan A. and Steber, Amanda L. and Muckle, Matt T. and Loomis, Ryan A. and Corby, Joanna F. and Martinez, Oscar and Crabtree, Kyle N. and Jewell, Philip R. and Hollis, Jan M. and Lovas, Frank J. and Vasquez, David and Nyiramahirwe, Jolie and Sciortino, Nicole and Johnson, Kennedy and McCarthy, Michael C. and Remijan, Anthony J. and Pate, Brooks H.},
  year = {2013},
  month = feb,
  journal = {The Astrophysical Journal},
  volume = {765},
  number = {1},
  pages = {L10},
  issn = {2041-8205, 2041-8213},
  doi = {10.1088/2041-8205/765/1/L10},
  urldate = {2025-05-21},
}

@article{HCCCHNH,
  title = {Propargylimine in the Laboratory and in Space: Millimetre-Wave Spectroscopy and Its First Detection in the {{ISM}}},
  shorttitle = {Propargylimine in the Laboratory and in Space},
  author = {Bizzocchi, L. and Prudenzano, D. and Rivilla, V. M. and {Pietropolli-Charmet}, A. and Giuliano, B. M. and Caselli, P. and {Mart{\'i}n-Pintado}, J. and {Jim{\'e}nez-Serra}, I. and Mart{\'i}n, S. and {Requena-Torres}, M. A. and {Rico-Villas}, F. and Zeng, S. and Guillemin, J.-C.},
  year = {2020},
  month = aug,
  journal = {Astronomy \& Astrophysics},
  volume = {640},
  pages = {A98},
  issn = {0004-6361, 1432-0746},
  doi = {10.1051/0004-6361/202038083},
  urldate = {2025-05-21},
}

@article{HNCNH,
  title = {{{INTERSTELLAR CARBODIIMIDE}} ({{HNCNH}}): {{A NEW ASTRONOMICAL DETECTION FROM THE GBT PRIMOS SURVEY VIA MASER EMISSION FEATURES}}},
  shorttitle = {{{INTERSTELLAR CARBODIIMIDE}} ({{HNCNH}})},
  author = {McGuire, Brett A. and Loomis, Ryan A. and Charness, Cameron M. and Corby, Joanna F. and Blake, Geoffrey A. and Hollis, Jan M. and Lovas, Frank J. and Jewell, Philip R. and Remijan, Anthony J.},
  year = {2012},
  month = oct,
  journal = {The Astrophysical Journal},
  volume = {758},
  number = {2},
  pages = {L33},
  issn = {2041-8205, 2041-8213},
  doi = {10.1088/2041-8205/758/2/L33},
  urldate = {2025-05-21},
}

@article{Zcyanomethanimine,
  title = {Abundant {{Z-cyanomethanimine}} in the Interstellar Medium: Paving the Way to the Synthesis of Adenine},
  shorttitle = {Abundant {{Z-cyanomethanimine}} in the Interstellar Medium},
  author = {Rivilla, V M and {Mart{\'i}n-Pintado}, J and {Jim{\'e}nez-Serra}, I and Zeng, S and Mart{\'i}n, S and {Armijos-Abenda{\~n}o}, J and {Requena-Torres}, M A and Aladro, R and Riquelme, D},
  year = {2019},
  month = feb,
  journal = {Monthly Notices of the Royal Astronomical Society: Letters},
  volume = {483},
  number = {1},
  pages = {L114-L119},
  issn = {1745-3925, 1745-3933},
  doi = {10.1093/mnrasl/sly228},
  urldate = {2025-05-21},
}

@article{ketenimine,
  title = {Detection of {{Ketenimine}} ({{CH}}                    {\textsubscript{2}}                    {{CNH}}) in {{Sagittarius B2}}({{N}}) {{Hot Cores}}},
  author = {Lovas, F. J. and Hollis, J. M. and Remijan, Anthony J. and Jewell, P. R.},
  year = {2006},
  month = jul,
  journal = {The Astrophysical Journal},
  volume = {645},
  number = {2},
  pages = {L137-L140},
  publisher = {American Astronomical Society},
  issn = {0004-637X, 1538-4357},
  doi = {10.1086/506324},
  urldate = {2025-05-22},
}

@article{HNCO,
  title = {Interstellar Isocyanic Acid},
  author = {Snyder, Lewis E. and Buhl, David},
  year = {1972},
  month = nov,
  journal = {Astrophysical Journal},
  volume = {177},
  pages = {619},
  doi = {10.1086/151739}
}

@article{H2CN,
  title = {Detection of a New Interstellar Molecule, {{H2CN}}},
  author = {Ohishi, Masatoshi and McGonagle, Douglas and Irvine, William M. and Yamamoto, Satoshi and Saito, Shuji},
  year = {1994},
  month = may,
  journal = {The Astrophysical Journal},
  volume = {427},
  pages = {L51},
  issn = {0004-637X, 1538-4357},
  doi = {10.1086/187362},
  urldate = {2025-05-14},
  langid = {english},
}

@article{DA,
  title = {Unveiling the {{Ionic Diels}}--{{Alder Reactions}} within the {{Molecular Electron Density Theory}}},
  author = {Domingo, Luis R. and {R{\'i}os-Guti{\'e}rrez}, Mar and Aurell, Mar{\'i}a Jos{\'e}},
  year = {2021},
  month = jun,
  journal = {Molecules},
  volume = {26},
  number = {12},
  pages = {3638},
  publisher = {MDPI AG},
  issn = {1420-3049},
  doi = {10.3390/molecules26123638},
  urldate = {2025-05-22},
}

@article{phenylradical,
  title = {Novel {{Products}} from {{C}}{\textsubscript{6}} {{H}}{\textsubscript{5}} + {{C}}{\textsubscript{6}} {{H}}{\textsubscript{6}} /{{C}}{\textsubscript{6}} {{H}}{\textsubscript{5}} {{Reactions}}},
  author = {Shukla, Bikau and Tsuchiya, Kentaro and Koshi, Mitsuo},
  year = {2011},
  month = jun,
  journal = {The Journal of Physical Chemistry A},
  volume = {115},
  number = {21},
  pages = {5284--5293},
  issn = {1089-5639, 1520-5215},
  doi = {10.1021/jp201817n},
  urldate = {2025-05-13},
}

@article{cyanobiphenyl,
  title = {{{ROTATIONAL SPECTROSCOPY OF}} 2- {{AND}} 4-{{CYANOBIPHENYL}}},
  author = {Schlemmer, Stephan},
  year = {2023},
  month = jun,
  publisher = {International Symposium on Molecular Spectroscopy},
  doi = {10.15278/isms.2023.6937},
  urldate = {2025-05-23},
}

@article{GOTHAM,
  title = {Early {{Science}} from {{GOTHAM}}: {{Project Overview}}, {{Methods}}, and the {{Detection}} of {{Interstellar Propargyl Cyanide}} ({{HCCCH2CN}}) in {{TMC-1}}},
  shorttitle = {Early {{Science}} from {{GOTHAM}}},
  author = {McGuire, Brett A. and Burkhardt, Andrew M. and Loomis, Ryan A. and Shingledecker, Christopher N. and Kelvin Lee, Kin Long and Charnley, Steven B. and Cordiner, Martin A. and Herbst, Eric and Kalenskii, Sergei and Momjian, Emmanuel and Willis, Eric R. and Xue, Ci and Remijan, Anthony J. and McCarthy, Michael C.},
  year = {2020},
  month = sep,
  journal = {The Astrophysical Journal Letters},
  volume = {900},
  number = {1},
  pages = {L10},
  publisher = {The American Astronomical Society},
  issn = {2041-8205},
  doi = {10.3847/2041-8213/aba632},
  urldate = {2025-05-23},
}

@article{Ecyanobutadiene,
  title = {Detection of {{Interstellar E-1-cyano-1}},3-Butadiene in {{GOTHAM Observations}} of {{TMC-1}}},
  author = {Cooke, Ilsa R. and Xue, Ci and Changala, P. Bryan and Shay, Hannah Toru and Byrne, Alex N. and Tang, Qi Yu and Fried, Zachary T. P. and Kelvin Lee, Kin Long and Loomis, Ryan A. and Lamberts, Thanja and Remijan, Anthony and Burkhardt, Andrew M. and Herbst, Eric and McCarthy, Michael C. and McGuire, Brett A.},
  year = {2023},
  month = may,
  journal = {The Astrophysical Journal},
  volume = {948},
  number = {2},
  pages = {133},
  publisher = {The American Astronomical Society},
  issn = {0004-637X},
  doi = {10.3847/1538-4357/acc584},
  urldate = {2025-05-23},
}

@article{Heterocycles,
  title = {A {{Search}} for {{Heterocycles}} in {{GOTHAM Observations}} of {{TMC-1}}},
  author = {Barnum, Timothy J. and Siebert, Mark A. and Lee, Kin Long Kelvin and Loomis, Ryan A. and Changala, P. Bryan and Charnley, Steven B. and Sita, Madelyn L. and Xue, Ci and Remijan, Anthony J. and Burkhardt, Andrew M. and McGuire, Brett A. and Cooke, Ilsa R.},
  year = {2022},
  month = may,
  journal = {The Journal of Physical Chemistry A},
  volume = {126},
  number = {17},
  pages = {2716--2728},
  publisher = {American Chemical Society},
  issn = {1089-5639},
  doi = {10.1021/acs.jpca.2c01435},
  urldate = {2025-05-23},
}

@article{stacking,
  title = {An Investigation of Spectral Line Stacking Techniques and Application to the Detection of {{HC11N}}},
  author = {Loomis, Ryan A. and Burkhardt, Andrew M. and Shingledecker, Christopher N. and Charnley, Steven B. and Cordiner, Martin A. and Herbst, Eric and Kalenskii, Sergei and Lee, Kin Long Kelvin and Willis, Eric R. and Xue, Ci and Remijan, Anthony J. and McCarthy, Michael C. and McGuire, Brett A.},
  year = {2021},
  month = jan,
  journal = {Nature Astronomy},
  volume = {5},
  number = {2},
  pages = {188--196},
  issn = {2397-3366},
  doi = {10.1038/s41550-020-01261-4},
  urldate = {2025-04-15},
  langid = {english},
}

@article{2propanimine,
  title = {Millimeter-Wave Spectrum of 2-Propanimine},
  author = {Zou, Luyao and Guillemin, Jean-Claude and Belloche, Arnaud and J{\o}rgensen, Jes K and Margul{\`e}s, Laurent and Motiyenko, Roman A and Groner, Peter},
  year = {2023},
  month = feb,
  journal = {Monthly Notices of the Royal Astronomical Society},
  volume = {520},
  number = {3},
  pages = {4089--4102},
  issn = {0035-8711, 1365-2966},
  doi = {10.1093/mnras/stad405},
  urldate = {2025-06-02},
}

@article{EZratio,
  title = {The {{Origin}} of the {{E}}/{{Z Isomer Ratio}} of {{Imines}} in the {{Interstellar Medium}}},
  author = {{Garc{\'i}a de la Concepci{\'o}n}, Juan and {Jim{\'e}nez-Serra}, Izaskun and Carlos Corchado, Jos{\'e} and Rivilla, V{\'i}ctor M. and {Mart{\'i}n-Pintado}, Jes{\'u}s},
  year = {2021},
  month = apr,
  journal = {The Astrophysical Journal Letters},
  volume = {912},
  number = {1},
  pages = {L6},
  publisher = {The American Astronomical Society},
  issn = {2041-8205},
  doi = {10.3847/2041-8213/abf650},
  urldate = {2025-05-29},
}

@article{hydrogenation,
  title = {Hydrogenation of Solid Hydrogen Cyanide {{HCN}} and Methanimine {{CH}}{\textsubscript{2}} {{NH}} at Low Temperature},
  author = {Theule, P. and Borget, F. and Mispelaer, F. and Danger, G. and Duvernay, F. and Guillemin, J. C. and Chiavassa, T.},
  year = {2011},
  month = oct,
  journal = {Astronomy \& Astrophysics},
  volume = {534},
  pages = {A64},
  issn = {0004-6361, 1432-0746},
  doi = {10.1051/0004-6361/201117494},
  urldate = {2025-06-02},
}

@article{Methaniminepath,
  title = {Low-{{Temperature Gas-Phase Formation}} of {{Methanimine}} ({{CH2NH}}; {{X1A}}{$\prime$})-the {{Simplest Imine}}-under {{Single-Collision Conditions}}},
  author = {Yang, Zhenghai and Medvedkov, Iakov A. and Goettl, Shane J. and Kaiser, Ralf I.},
  year = {2023},
  month = sep,
  journal = {The Journal of Physical Chemistry Letters},
  volume = {14},
  number = {38},
  pages = {8500--8506},
  publisher = {American Chemical Society},
  doi = {10.1021/acs.jpclett.3c02360},
  urldate = {2025-06-04},
}

@article{phenylmethanimine,
  title = {A {{Journey}} from {{Thermally Tunable Synthesis}} to {{Spectroscopy}} of {{Phenylmethanimine}} in {{Gas Phase}} and {{Solution}}},
  author = {Melli, Alessio and Potenti, Simone and Melosso, Mattia and Herbers, Sven and Spada, Lorenzo and Gualandi, Andrea and Lengsfeld, Kevin G. and Dore, Luca and Buschmann, Philipp and Cozzi, Pier Giorgio and Grabow, Jens-Uwe and Barone, Vincenzo and Puzzarini, Cristina},
  year = {2020},
  journal = {Chemistry -- A European Journal},
  volume = {26},
  number = {65},
  pages = {15016--15022},
  issn = {1521-3765},
  doi = {10.1002/chem.202003270},
  urldate = {2025-06-03},
}

@article{smallPAHs,
  title = {Rotational {{Spectra}} of {{Small PAHs}}: {{Acenaphthene}}, {{Acenaphthylene}}, {{Azulene}}, and {{Fluorene}}},
  shorttitle = {Rotational {{Spectra}} of {{Small PAHs}}},
  author = {Thorwirth, S. and Theule, P. and Gottlieb, C. A. and McCarthy, M. C. and Thaddeus, P.},
  year = {2007},
  month = jun,
  journal = {The Astrophysical Journal},
  volume = {662},
  number = {2},
  pages = {1309--1314},
  publisher = {American Astronomical Society},
  issn = {0004-637X, 1538-4357},
  doi = {10.1086/518026},
  urldate = {2025-07-10},
}

@article{H2CNCN,
  title = {First {{Detection}} in {{Space}} of the {{High-energy Isomer}} of {{Cyanomethanimine}}: {{H}}{\textsubscript{2}}{{CNCN}}},
  shorttitle = {First {{Detection}} in {{Space}} of the {{High-energy Isomer}} of {{Cyanomethanimine}}},
  author = {San Andr{\'e}s, David and Rivilla, V{\'i}ctor M. and Colzi, Laura and {Jim{\'e}nez-Serra}, Izaskun and {Mart{\'i}n-Pintado}, Jes{\'u}s and Meg{\'i}as, Andr{\'e}s and {L{\'o}pez-Gallifa}, {\'A}lvaro and {Mart{\'i}nez-Henares}, Antonio and Massalkhi, Sarah and Zeng, Shaoshan and {Sanz-Novo}, Miguel and Tercero, Bel{\'e}n and De Vicente, Pablo and Mart{\'i}n, Sergio and Requena Torres, Miguel Angel and Molpeceres, Germ{\'a}n and Garc{\'i}a De La Concepci{\'o}n, Juan},
  year = {2024},
  month = may,
  journal = {The Astrophysical Journal},
  volume = {967},
  number = {1},
  pages = {39},
  publisher = {American Astronomical Society},
  issn = {0004-637X, 1538-4357},
  doi = {10.3847/1538-4357/ad3af3},
  urldate = {2025-07-10},
}

@article{H2NC,
  title = {Interstellar Detection of the Simplest Aminocarbyne {{H}}{\textsubscript{2}}{{NC}}: An Ignored but Abundant Molecule},
  shorttitle = {Interstellar Detection of the Simplest Aminocarbyne {{H}}{\textsubscript{2}}{{NC}}},
  author = {Cabezas, C. and Ag{\'u}ndez, M. and Marcelino, N. and Tercero, B. and Cuadrado, S. and Cernicharo, J.},
  year = {2021},
  month = oct,
  journal = {Astronomy \& Astrophysics},
  volume = {654},
  pages = {A45},
  publisher = {EDP Sciences},
  issn = {0004-6361, 1432-0746},
  doi = {10.1051/0004-6361/202141491},
  urldate = {2025-07-10},
}

@article{DR5.3,
  title = {The {{Molecular Inventory}} of {{TMC-1}} with {{GOTHAM Observations}}},
  author = {Xue, Ci and Byrne, Alex N. and Morgan, Larry and Wenzel, Gabi and Changala, P. Bryan and Fried, Zachary T. P. and Loomis, Ryan A. and Remijan, Anthony and Bergin, Edwin A. and Cooke, Ilsa R. and Frayer, David and Burkhardt, Andrew M. and Charnley, Steven B. and Cordiner, Martin A. and Lipnicky, Andrew and McCarthy, Michael C. and McGuire, Brett A.},
  year = 2025,
  month = nov,
  journal = {The Astrophysical Journal Supplement Series},
  volume = {281},
  number = {1},
  pages = {9},
  issn = {0067-0049, 1538-4365},
  doi = {10.3847/1538-4365/ae04e5},
  urldate = {2025-11-20},
  langid = {english}
}

@misc{g16c02,
  title = {Gaussian 16 {{Revision C}}.02},
  author = {Frisch, M. J. and Trucks, G. W. and Schlegel, H. B. and Scuseria, G. E. and Robb, M. A. and Cheeseman, J. R. and Scalmani, G. and Barone, V. and Petersson, G. A. and Nakatsuji, H. and Li, X. and Caricato, M. and Marenich, A. V. and Bloino, J. and Janesko, B. G. and Gomperts, R. and Mennucci, B. and Hratchian, H. P. and Ortiz, J. V. and Izmaylov, A. F. and Sonnenberg, J. L. and {Williams-Young}, D. and Ding, F. and Lipparini, F. and Egidi, F. and Goings, J. and Peng, B. and Petrone, A. and Henderson, T. and Ranasinghe, D. and Zakrzewski, V. G. and Gao, J. and Rega, N. and Zheng, G. and Liang, W. and Hada, M. and Ehara, M. and Toyota, K. and Fukuda, R. and Hasegawa, J. and Ishida, M. and Nakajima, T. and Honda, Y. and Kitao, O. and Nakai, H. and Vreven, T. and Throssell, K. and Montgomery, Jr., J. A. and Peralta, J. E. and Ogliaro, F. and Bearpark, M. J. and Heyd, J. J. and Brothers, E. N. and Kudin, K. N. and Staroverov, V. N. and Keith, T. A. and Kobayashi, R. and Normand, J. and Raghavachari, K. and Rendell, A. P. and Burant, J. C. and Iyengar, S. S. and Tomasi, J. and Cossi, M. and Millam, J. M. and Klene, M. and Adamo, C. and Cammi, R. and Ochterski, J. W. and Martin, R. L. and Morokuma, K. and Farkas, O. and Foresman, J. B. and Fox, D. J.},
  year = {2019}
}

@article{orca6.0,
  title = {Software {{Update}}: {{The}} {{{\textsc{ORCA}}}} {{Program System}}---{{Version}} 6.0},
  shorttitle = {Software {{Update}}},
  author = {Neese, Frank},
  year = {2025},
  month = mar,
  journal = {WIREs Computational Molecular Science},
  volume = {15},
  number = {2},
  publisher = {Wiley},
  issn = {1759-0876, 1759-0884},
  doi = {10.1002/wcms.70019},
  urldate = {2025-07-16},
}

@article{orca2020,
  title = {The {{ORCA}} Quantum Chemistry Program Package},
  author = {Neese, Frank and Wennmohs, Frank and Becker, Ute and Riplinger, Christoph},
  year = {2020},
  month = jun,
  journal = {The Journal of Chemical Physics},
  volume = {152},
  number = {22},
  publisher = {AIP Publishing},
  issn = {0021-9606, 1089-7690},
  doi = {10.1063/5.0004608},
  urldate = {2025-07-16},
}

@article{phenyl,
  title = {Hyperfine-{{Resolved Rotational Spectroscopy}} of {{Phenyl Radical}}},
  author = {Changala, P. Bryan and McCarthy, Michael C.},
  year = {2023},
  month = jun,
  journal = {The Journal of Physical Chemistry Letters},
  volume = {14},
  number = {23},
  pages = {5370--5376},
  publisher = {American Chemical Society},
  doi = {10.1021/acs.jpclett.3c01243},
  urldate = {2025-07-24},
}

@article{gasphasekinetics,
  title = {Gas-Phase {{Chemistry}} in the {{Interstellar Medium}}: {{The Role}} of {{Laboratory Astrochemistry}}},
  shorttitle = {Gas-Phase {{Chemistry}} in the {{Interstellar Medium}}},
  author = {Puzzarini, Cristina},
  year = {2022},
  month = feb,
  journal = {Frontiers in Astronomy and Space Sciences},
  volume = {8},
  publisher = {Frontiers},
  issn = {2296-987X},
  doi = {10.3389/fspas.2021.811342},
  urldate = {2025-07-25},
}

@article{addCN,
  title = {The Reactivity of Pyridine in Cold Interstellar Environments: {{The}} Reaction of Pyridine with the {{CN}} Radical},
  shorttitle = {The Reactivity of Pyridine in Cold Interstellar Environments},
  author = {Heitk{\"a}mper, Juliane and Suchaneck, Sarah and {Garc{\'i}a de la Concepci{\'o}n}, Juan and K{\"a}stner, Johannes and Molpeceres, Germ{\'a}n},
  year = {2022},
  month = dec,
  journal = {Frontiers in Astronomy and Space Sciences},
  volume = {9},
  publisher = {Frontiers},
  issn = {2296-987X},
  doi = {10.3389/fspas.2022.1020635},
  urldate = {2025-07-25},
}

@article{anharmonic,
  title = {Accurate {{Harmonic}}/{{Anharmonic Vibrational Frequencies}} for {{Open-Shell Systems}}: {{Performances}} of the {{B3LYP}}/{{N07D Model}} for {{Semirigid Free Radicals Benchmarked}} by {{CCSD}}({{T}}) {{Computations}}},
  shorttitle = {Accurate {{Harmonic}}/{{Anharmonic Vibrational Frequencies}} for {{Open-Shell Systems}}},
  author = {Puzzarini, Cristina and Biczysko, Malgorzata and Barone, Vincenzo},
  year = {2010},
  month = mar,
  journal = {Journal of Chemical Theory and Computation},
  volume = {6},
  number = {3},
  pages = {828--838},
  publisher = {American Chemical Society},
  issn = {1549-9618},
  doi = {10.1021/ct900594h},
  urldate = {2025-07-25},
}

@article{KiSThelP,
  title = {{{KiSThelP}}: {{A}} Program to Predict Thermodynamic Properties and Rate Constants from Quantum Chemistry Results{\dag}},
  shorttitle = {{{KiSThelP}}},
  author = {Canneaux, S{\'e}bastien and Bohr, Fr{\'e}d{\'e}ric and Henon, Eric},
  year = {2014},
  journal = {Journal of Computational Chemistry},
  volume = {35},
  number = {1},
  pages = {82--93},
  issn = {1096-987X},
  doi = {10.1002/jcc.23470},
  urldate = {2025-08-12},
  copyright = {Copyright {\copyright} 2013 Wiley Periodicals, Inc.},
  langid = {english},
}

@article{D3,
  title = {A Consistent and Accurate Ab Initio Parametrization of Density Functional Dispersion Correction ({{DFT-D}}) for the 94 Elements {{H-Pu}}},
  author = {Grimme, Stefan and Antony, Jens and Ehrlich, Stephan and Krieg, Helge},
  year = {2010},
  month = apr,
  journal = {The Journal of Chemical Physics},
  volume = {132},
  number = {15},
  pages = {154104},
  issn = {0021-9606},
  doi = {10.1063/1.3382344},
  urldate = {2025-08-12},
}

@article{def2TZVPP1,
  title = {Accurate {{Coulomb-fitting}} Basis Sets for {{H}} to {{Rn}}},
  author = {Weigend, Florian},
  year = {2006},
  month = feb,
  journal = {Physical Chemistry Chemical Physics},
  volume = {8},
  number = {9},
  pages = {1057--1065},
  publisher = {The Royal Society of Chemistry},
  issn = {1463-9084},
  doi = {10.1039/B515623H},
  urldate = {2025-08-12},
  langid = {english},
}

@article{def2TZVPP2,
  title = {Balanced Basis Sets of Split Valence, Triple Zeta Valence and Quadruple Zeta Valence Quality for {{H}} to {{Rn}}: {{Design}} and Assessment of Accuracy},
  shorttitle = {Balanced Basis Sets of Split Valence, Triple Zeta Valence and Quadruple Zeta Valence Quality for {{H}} to {{Rn}}},
  author = {Weigend, Florian and Ahlrichs, Reinhart},
  year = {2005},
  month = aug,
  journal = {Physical Chemistry Chemical Physics},
  volume = {7},
  number = {18},
  pages = {3297--3305},
  publisher = {The Royal Society of Chemistry},
  issn = {1463-9084},
  doi = {10.1039/B508541A},
  urldate = {2025-08-12},
  langid = {english},
}

@article{M062X,
  title = {The {{M06}} Suite of Density Functionals for Main Group Thermochemistry, Thermochemical Kinetics, Noncovalent Interactions, Excited States, and Transition Elements: Two New Functionals and Systematic Testing of Four {{M06-class}} Functionals and 12 Other Functionals},
  shorttitle = {The {{M06}} Suite of Density Functionals for Main Group Thermochemistry, Thermochemical Kinetics, Noncovalent Interactions, Excited States, and Transition Elements},
  author = {Zhao, Yan and Truhlar, Donald G.},
  year = {2008},
  journal= {Theoretical Chemistry Accounts},
  volume = {120},
  number = {1},
  pages = {215--241},
  issn = {1432-2234},
  doi = {10.1007/s00214-007-0310-x},
  langid = {english}
}

@article{6-311g,
  title = {Self-consistent Molecular Orbital Methods. {{XX}}. {{A}} Basis Set for Correlated Wave Functions},
  author = {Krishnan, R. and Binkley, J. S. and Seeger, R. and Pople, J. A.},
  year = {1980},
  month = jan,
  journal = {The Journal of Chemical Physics},
  volume = {72},
  number = {1},
  pages = {650--654},
  issn = {0021-9606},
  doi = {10.1063/1.438955},
  urldate = {2025-08-12}
}

@article{B3LYP,
  title = {Density-functional Thermochemistry. {{III}}. {{The}} Role of Exact Exchange},
  author = {Becke, Axel D.},
  year = {1993},
  month = apr,
  journal = {The Journal of Chemical Physics},
  volume = {98},
  number = {7},
  pages = {5648--5652},
  issn = {0021-9606},
  doi = {10.1063/1.464913},
  urldate = {2025-08-12}
}

@article{D3BJ,
  title = {Effect of the Damping Function in Dispersion Corrected Density Functional Theory},
  author = {Grimme, Stefan and Ehrlich, Stephan and Goerigk, Lars},
  year = {2011},
  journal = {Journal of Computational Chemistry},
  volume = {32},
  number = {7},
  pages = {1456--1465},
  issn = {1096-987X},
  doi = {10.1002/jcc.21759},
  urldate = {2025-08-12},
  copyright = {Copyright {\copyright} 2011 Wiley Periodicals, Inc.},
  langid = {english}
}

@article{VPT2,
  title = {General {{Perturbative Approach}} for {{Spectroscopy}}, {{Thermodynamics}}, and {{Kinetics}}: {{Methodological Background}} and {{Benchmark Studies}}},
  shorttitle = {General {{Perturbative Approach}} for {{Spectroscopy}}, {{Thermodynamics}}, and {{Kinetics}}},
  author = {Bloino, Julien and Biczysko, Malgorzata and Barone, Vincenzo},
  year = {2012},
  month = mar,
  journal = {Journal of Chemical Theory and Computation},
  volume = {8},
  number = {3},
  pages = {1015--1036},
  publisher = {American Chemical Society},
  issn = {1549-9618},
  doi = {10.1021/ct200814m},
  urldate = {2025-08-19}
}

@article{augccpvtz/c,
  title = {Efficient Use of the Correlation Consistent Basis Sets in Resolution of the Identity {{MP2}} Calculations},
  author = {Weigend, Florian and Köhn, Andreas and Hättig, Christof},
  year = {2002},
  date = {2002-02-22},
  journal = {The Journal of Chemical Physics},
  shortjournal = {J. Chem. Phys.},
  volume = {116},
  number = {8},
  pages = {3175--3183},
  issn = {0021-9606},
  doi = {10.1063/1.1445115}
}

@article{augccpvtz,
  title = {Electron Affinities of the First-row Atoms Revisited. {{Systematic}} Basis Sets and Wave Functions},
  author = {Kendall, Rick A. and Dunning, Jr., Thom H. and Harrison, Robert J.},
  year = {1992},
  month = may,
  journal = {The Journal of Chemical Physics},
  volume = {96},
  number = {9},
  pages = {6796--6806},
  issn = {0021-9606},
  doi = {10.1063/1.462569},
  urldate = {2025-08-19}
}

@article{DLPNO,
  type = {Journal {{Article}}},
  title = {Communication: {{An}} Improved Linear Scaling Perturbative Triples Correction for the Domain Based Local Pair-Natural Orbital Based Singles and Doubles Coupled Cluster Method {{DLPNO-CCSD}}({{T}})},
  author = {Guo, Y. and Riplinger, C. and Becker, U. and Liakos, D.G. and Minenkov, Y. and Cavallo, L. and Neese, F.},
  year = {2018},
  journal = {Journal of Chemical Physics},
  volume = {148},
  pages = {5},
  doi = {10.1063/1.5011798}
}

@article{ROHF,
  title = {Analytic Energy Gradients for Open-Shell Coupled-Cluster Singles and Doubles ({{CCSD}}) Calculations Using Restricted Open-Shell {{Hartree}}---{{Fock}} ({{ROHF}}) Reference Functions},
  author = {Gauss, J{\"u}rgen and Lauderdale, Walter J. and Stanton, John F. and Watts, John D. and Bartlett, Rodney J.},
  year = {1991},
  month = aug,
  journal = {Chemical Physics Letters},
  volume = {182},
  number = {3},
  pages = {207--215},
  issn = {0009-2614},
  doi = {10.1016/0009-2614(91)80203-A},
  urldate = {2025-08-19}
}

@article{tunneling,
  title = {Tunneling Corrections to Unimolecular Rate Constants, with Application to Formaldehyde},
  author = {Miller, William H.},
  year = {1979},
  month = nov,
  journal = {Journal of the American Chemical Society},
  volume = {101},
  number = {23},
  pages = {6810--6814},
  publisher = {American Chemical Society},
  issn = {0002-7863},
  doi = {10.1021/ja00517a004},
  urldate = {2024-07-29}
}

@article{cC3H2S,
  title = {The {{Missing Link}} of {{Sulfur Chemistry}} in {{TMC-1}}: {{The Detection}} of {\emph{c}} -{{C}}{\textsubscript{3}} {{H}}{\textsubscript{2}} {{S}} from the {{GOTHAM Survey}}},
  shorttitle = {The {{Missing Link}} of {{Sulfur Chemistry}} in {{TMC-1}}},
  author = {Remijan, Anthony J. and Changala, P. Bryan and Xue, Ci and Yuan, Elsa Q. H. and Duffy, Miya and Scolati, Haley N. and Shingledecker, Christopher N. and Speak, Thomas H. and Cooke, Ilsa R. and Loomis, Ryan and Burkhardt, Andrew M. and Fried, Zachary T. P. and Wenzel, Gabi and Lipnicky, Andrew and McCarthy, Michael C. and McGuire, Brett A.},
  year = {2025},
  month = apr,
  journal = {The Astrophysical Journal},
  volume = {982},
  number = {2},
  pages = {191},
  issn = {0004-637X, 1538-4357},
  doi = {10.3847/1538-4357/adb84c},
  urldate = {2025-08-27},
  langid = {english}
}

@article{DLPNOCCSDT,
  title = {Comparison of the Hydrogen Extraction Reactions of Isopentane Molecules and Ions},
  author = {Gao, Yi and Yang, Bin and Zhao, Yang},
  year = {2022},
  journal = {Physical Chemistry Chemical Physics},
  volume = {24},
  number = {39},
  pages = {24524--24534},
  publisher = {Royal Society of Chemistry},
  doi = {10.1039/D2CP01047J},
  urldate = {2025-08-27},
  langid = {english}
}

@article{phenalene,
  title = {Discovery of Interstellar Phenalene (c-{{C13H10}}): {{A}} New Piece in the Chemical Puzzle of {{PAHs}} in Space},
  shorttitle = {Discovery of Interstellar Phenalene (c-{{C13H10}})},
  author = {Cabezas, C. and Ag{\'u}ndez, M. and P{\'e}rez, C. and {Villar-Castro}, D. and Molpeceres, G. and P{\'e}rez, D. and Steber, A. L. and Fuentetaja, R. and Tercero, B. and Marcelino, N. and Lesarri, A. and de Vicente, P. and Cernicharo, J.},
  year = 2025,
  month = sep,
  journal = {Astronomy \& Astrophysics},
  volume = {701},
  pages = {L8},
  publisher = {EDP Sciences},
  issn = {0004-6361, 1432-0746},
  doi = {10.1051/0004-6361/202556687},
  urldate = {2025-11-03},
  copyright = {\copyright{} The Authors 2025},
  langid = {english}
}

@article{MRR,
  title = {Analysis of Isomeric Mixtures by Molecular Rotational Resonance Spectroscopy},
  author = {Neill, Justin L. and Evangelisti, Luca and Pate, Brooks H.},
  year = 2023,
  journal = {Analytical Science Advances},
  volume = {4},
  number = {5-6},
  pages = {204--219},
  issn = {2628-5452},
  doi = {10.1002/ansa.202300021},
  urldate = {2025-11-21},
  copyright = {\copyright{} 2023 The Authors. Analytical Science Advances published by Wiley-VCH GmbH.},
  langid = {english}
}
\bibliographystyle{aasjournal}

\end{document}